\documentclass{article}
\usepackage{amsmath,amssymb,amsfonts}
\usepackage{algorithmic}
\usepackage{graphicx}
\usepackage{textcomp}
\usepackage{xcolor}
\usepackage{hyperref}
\usepackage{physics}
\usepackage[]{algorithm2e}
\usepackage{mdframed}
\usepackage{caption}
\usepackage{subcaption}
\usepackage{flafter}
\usepackage{lmodern}
\usepackage{float}
\usepackage{physics}
\usepackage{longfbox}
\usepackage{tabularx}
\usepackage{appendix}
\usepackage{pgfplots}
\makeatletter
\newdimen\@tempdimd
\makeatother
\newmdtheoremenv{heur}{Heuristic}
\usepackage[T1]{fontenc}
\fboxset{rounded, border-width=0.2pt, padding={0.2ex,0.4ex}}
\usepackage{tikz}
\usetikzlibrary{decorations.pathreplacing}
\usetikzlibrary{patterns}
\usepackage[english]{babel}
\usetikzlibrary {arrows.meta} 
\usepackage{authblk}

\title{\LARGE Assessing the Efficacy of Heuristic-Based Address Clustering for Bitcoin}

\author[1,2]{Hugo Schnoering\thanks{hschnoering@coinshares.com}}
\author[1]{Pierre Porthaux\thanks{pporthaux@coinshares.com}}
\author[2]{Michalis Vazirgiannis\thanks{mvazirg@lix.polytechnique.fr}}

\affil[1]{\small Coinshares}
\affil[2]{\small Ecole Polytechnique}

\begin{document}

\maketitle

\abstract{

Exploring transactions within the Bitcoin blockchain entails examining the transfer of bitcoins among several hundred million entities. However, it is often impractical and resource-consuming to study such a vast number of entities. Consequently, entity clustering serves as an initial step in most analytical studies. This process often employs heuristics grounded in the practices and behaviors of these entities. In this research, we delve into the examination of two widely used heuristics, alongside the introduction of four novel ones. Our contribution includes the introduction of the \textit{clustering ratio}, a metric designed to quantify the reduction in the number of entities achieved by a given heuristic. The assessment of this reduction ratio plays an important role in justifying the selection of a specific heuristic for analytical purposes. Given the dynamic nature of the Bitcoin system, characterized by a continuous increase in the number of entities on the blockchain, and the evolving behaviors of these entities, we extend our study to explore the temporal evolution of the clustering ratio for each heuristic. This temporal analysis enhances our understanding of the effectiveness of these heuristics over time.

}

\section*{Introduction}

Bitcoin \cite{nakamoto2008bitcoin} is a peer-to-peer network designed to transfer value in the form of bitcoins between participants. These value transfers are encoded in transactions, and the entire sequence of transactions since the network's inception is chronicled on a public and decentralized ledger known as the Bitcoin blockchain. Each participant, or \textit{user}, can choose to create an arbitrarily large number of identities, or \textit{addresses}, to engage with the network. As a result, studying Bitcoin transactions from its inception to date involves examining transfers of bitcoins between several hundred million different addresses. Several heuristics have been developed to detect addresses that may belong to the same user. Aggregating large numbers of addresses reduces the total number of entities, and will thus facilitate the analysis of the Bitcoin blockchain.

\paragraph{Related Works} The explosion in the number of Bitcoin addresses has been observed in several studies. The issue of reducing the number of entities has thus been raised multiple times \cite{filtz2017evolution,harlev2018breaking}. In order to address this concern and make the analysis of the Bitcoin network and blockchain feasible, several authors have employed an initial clustering step, i.e., creating super-entities by merging entities based on specific rules called heuristics. These heuristics rely on the micro-structure of transactions, key management patterns, and Unspent Transaction Outputs (UTXO), aiming to identify identities/pseudonyms that could reasonably belong to the same user.\\ 

The most commonly used heuristics are the \textit{common-input-ownership heuristic} and the \textit{change heuristic}. The common-input-ownership heuristic was first mentioned in the Bitcoin white paper authored by the pseudonym Nakamoto. This heuristic was further developed in the works of \cite{androulaki2013evaluating}, \cite{reid2013analysis}, and \cite{meiklejohn2013fistful}, all agreeing on the same definition of this heuristic. On the other hand, the change heuristic exists in various forms, with over twenty identified according to Moser  \textit{et al.} \cite{moser2022resurrecting}. The initial versions emerged in 2013 in the works of \cite{androulaki2013evaluating} and \cite{meiklejohn2013fistful}. The heuristic described in the former paper focuses on transactions with two recipients, whereas the latter defines it potentially involving more than two recipients. These change heuristics have since undergone modifications or refinements \cite{ermilov2017automatic, harrigan2016unreasonable}. Other less common heuristics have been developed, specifically targeting Coinbase transactions or transactions involving mining pools \cite{he2022bitcoin}. \\

Several studies initially employ these heuristics to reduce the number of entities. Subsequently, traditional machine learning algorithms are applied to further decrease the entity count. Androulaki et al. \cite{androulaki2013evaluating} utilized K-means and Hierarchical Clustering algorithms on a node network obtained through the application of specific heuristics. Remy et al. \cite{remy2018tracking} constructed a network where nodes represent clusters discovered by the common-input-ownership heuristic, adding edges from senders to recipients under specific conditions. The Louvain algorithm is subsequently applied to this constructed network to identify the final entities. He et al. \cite{he2022bitcoin} also use the Louvain algorithm on a cluster graph obtained through the application of four different heuristics. Chang et al. \cite{chang2018improving} used common heuristics and continue clustering by detecting certain predefined transaction patterns and applying specific clustering rules. Moser et al. \cite{moser2022resurrecting} trained a Random Forest classifier that, based on conclusions from various heuristics and transaction features, detects the change address in a transaction. \\ 

The assessment of the relevance of heuristics, i.e., their actual ability to group identities belonging to the same user, has been addressed in several studies \cite{meiklejohn2013fistful,chang2018improving,moser2022resurrecting}. However, the absence of a ground truth dataset representing a significant number of users of different types makes the task challenging. Hence, the utilization of a particular heuristic primarily relies on conviction. Conversely, measuring the effectiveness of reducing entities through a heuristic, i.e., its ability to decrease the number of entities, is more direct. Zhang et al. \cite{zhang2020heuristic} introduced in their work the address reduction ratio and evaluated its value for various combinations of heuristics. They also highlighted the additional contribution to address reduction by two change heuristics compared to the common-input-ownership heuristic. Finally, several studies \cite{harrigan2016unreasonable,moser2022resurrecting} have focused on the dynamics of cluster formation over time, as well as the issue of cluster collapse.

\paragraph{Contributions} Our contributions are manifold. In section \ref{sec:concepts}, we introduce various concepts related to Bitcoin and define notations that allow for the precise definition of the clustering heuristics. In section \ref{sec:heuristics}, we examine six clustering heuristics, including the widely used common-input-ownership heuristic, and our own version of the change address heuristic. We also design four novel heuristics:
\begin{enumerate}
\item A refined version of the common-input-ownership heuristic that excludes entities involved in CoinJoin transactions from being merged.
\item A first alternative change address heuristic that detects change amounts, leveraging a known human bias for round amounts.
\item A second alternative change address heuristic that identifies change addresses by assuming input UTXOs are chosen to minimize links between the addresses belonging to the same user.
\item A third version of the common-input-ownership heuristic aimed at identifying consolidation transactions conducted by businesses or services using Bitcoin.
\end{enumerate}
Finally, in section \ref{sec:results}, we introduce the clustering ratio as a measure of a heuristic's clustering effectiveness. We proceed to evaluate the clustering effectiveness of the designed heuristics and also assess the effectiveness of a combination of four heuristics. To measure the temporal effectiveness of these heuristics, we calculate these ratios across different block indices, extending our analysis up to block index 700,000.

\section{Address Clustering}
\label{sec:concepts}

\subsection{Users and Keys} 

Bitcoin relies on the principles of public-key cryptography. A user  $u$ can possess or control a set of one or multiple private keys, represented collectively as $\mathcal{K}_u$. Each private key $k \in \mathcal{K}_u$ secures a portion of $u$'s wealth. Thus, $u$ must  keep its private keys secret, as their knowledge allows to access and control the associated funds. Instead of exposing the private keys to the network, Bitcoin employs a range of derived identifiers, such as public keys, public key hashes, and various other quantities. All of these identifiers are computed from private keys through one-way cryptographic / hash functions. These pseudonymous representations of $u$, collectively referred to as \textit{addresses}, can thus be safely shared with others, enabling the other participants to identify $u$. The generation of a private key by a user is a relatively straightforward process, involving the selection of a number from the set $\{0, 1\}^{256}$ without the need for intermediaries. This allows a user to potentially use a very large number of keys, and therefore have as many identities on the network. However, managing multiple keys can be complex, especially concerning their storage or the creation of cryptographic signatures required for transactions. The introduction of Hierarchical Deterministic (HD) wallets, as outlined in Bitcoin Improvement Proposal 32 \cite{bip32_wuille}, has made private key management simpler. Hierarchical deterministic wallets rely on a master key from which new private keys can be deterministically derived. This improvement simplifies the creation and the storage of private keys, enabling a user $u$ to store his funds in different keys that cannot be easily linked by observers.  This enhances the overall privacy and confidentiality of his Bitcoin transactions.

\subsection{TXO and Transactions} 

\paragraph{Transaction Output (TXO)} The unit of value of the Bitcoin network, the bitcoin, exists in the form of \textit{transaction output}. A \textit{transaction output} (TXO) $\tau$  is defined by two components: a value $v \in \mathbb{N}$ in satoshis (1 satoshi = $10^{-8}$ bitcoin) and $\ket{p}$, a part of a computer program written in Bitcoin script. $\ket{p}$ is called a \textit{locking script} because  it specifies the conditions under which $\tau$,  and by consequent the associated value $v$, can be spent. In most cases, $\ket{p}$ specifies one or several identifiers derived from private keys $k_1, ..., k_i$ allowing only users who knows $k_1, ..., k_i$ to 'unlock' $\ket{p}$ and spend the output $\tau$. The group of users $u_1, ..., u_j$ that have the knowledge of these keys are considered the owners of $\tau$.  Since $\ket{p}$ is derived from private keys, it is also referred to as an address.

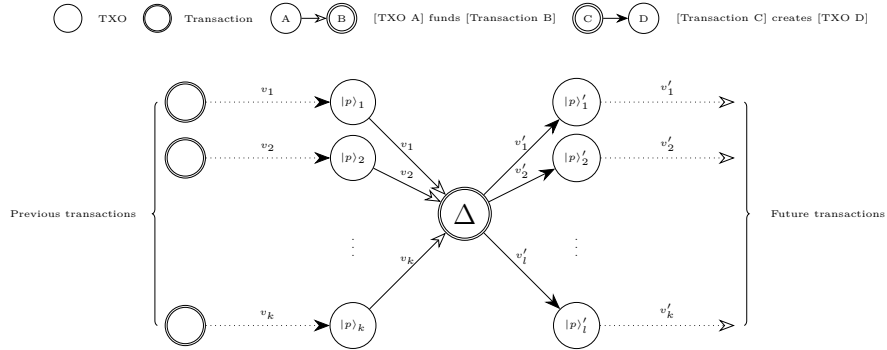
\begin{figure}[H]
\centering

\resizebox{0.99\textwidth}{!}{
\begin{tikzpicture}

\node[circle, draw, font=\tiny, minimum size=0.50cm] (A) at (-7.1, 3.5) {};
\node[font=\tiny] (A) at (-6.3, 3.5) {TXO};

\node[circle, draw, font=\tiny, minimum size=0.45cm] (A) at (-5.5, 3.5) {};
\node[circle, draw, font=\tiny, minimum size=0.50cm] (A) at (-5.5, 3.5) {};
\node[font=\tiny] (A) at (-4.4, 3.5) {Transaction};

\node[circle, draw, font=\tiny, minimum size=0.50cm] (TXOA) at (-3.2, 3.5) {A};
\node[circle, draw, font=\tiny, minimum size=0.50cm] (transationB) at (-2.2, 3.5) {B};
\node[circle, draw, font=\tiny, minimum size=0.45cm] (A) at (-2.2, 3.5) {};
\draw[arrows = {-Stealth[length=6pt,fill=none]}] (TXOA) -- (transationB);

\node[font=\tiny] (A) at (-0., 3.5) {[TXO A] funds [Transaction B]};

\node[circle, draw, font=\tiny, minimum size=0.50cm] (transationC) at (2.2, 3.5) {C};
\node[circle, draw, font=\tiny, minimum size=0.45cm] (A) at (2.2, 3.5) {};
\node[circle, draw, font=\tiny, minimum size=0.50cm] (TXOD) at (3.2, 3.5) {D};
\draw[arrows = {-Stealth[length=6pt]}] (transationC) -- (TXOD);

\node[font=\tiny] (A) at (5.5, 3.5) {[Transaction C] creates [TXO D]};

\draw [decorate, decoration = {brace}] (-5.5,-2) --  (-5.5,2);
\draw [decorate, decoration = {brace}] (5,2) --  (5,-2);
\node[circle, font=\tiny] (prevtransaction) at (-7,0) {Previous transactions};
\node[circle, font=\tiny] (prevtransaction) at (6.5,0) {Future transactions};


\node[circle, draw, font=\tiny, minimum size=0.72cm] (transactionin1) at (-5,2) {};
\node[circle, draw, font=\tiny, minimum size=0.65cm] (transactionin1bis) at (-5,2) {};
\node[circle, draw, font=\tiny, minimum size=0.72cm] (transactionin2) at (-5,1) {};
\node[circle, draw, font=\tiny, minimum size=0.65cm] (transactionin2bis) at (-5,1) {};
\node[circle, draw, font=\tiny, minimum size=0.72cm] (transactionink) at (-5,-2) {};
\node[circle, draw, font=\tiny, minimum size=0.65cm] (transactioninibis) at (-5,-2) {};

\node[circle, draw, font=\tiny] (nodein1) at (-2,2) {$\ket{p}_1$};
\node[circle, draw, font=\tiny] (nodein2) at (-2,1) {$\ket{p}_2$};
\node[circle, font=\tiny] (nodeinj) at (-2,-0.5) {$\vdots$};
\node[circle, draw, font=\tiny] (nodeink) at (-2,-2) {$\ket{p}_k$};

\node[circle, draw, font=\Large, minimum size=0.95cm] (nodedelta) at (0,0) {};
\node[circle, draw, font=\Large, minimum size=0.8cm] (nodedelta1) at (0,0) {$\Delta$};

\node[circle, font=\tiny] (transactionout1) at (5,2) {};
\node[circle, font=\tiny] (transactionout2) at (5,1) {};
\node[circle, font=\tiny] (transactionoutk) at (5,-2) {};

\node[circle, draw, font=\tiny] (nodeout1) at (2,2) {$\ket{p}_1^\prime$};
\node[circle, draw, font=\tiny] (nodeout2) at (2,1) {$\ket{p}_2^\prime$};
\node[circle, font=\tiny] (nodeoutj) at (2,-0.5) {$\vdots$};
\node[circle, draw, font=\tiny] (nodeoutk) at (2,-2) {$\ket{p}_l^\prime$};


\draw[arrows = {-Stealth[length=8pt]}, dotted] (transactionin1) -- (nodein1) node[midway, above,font=\tiny] {$v_1$};
\draw[arrows = {-Stealth[length=8pt]}, dotted] (transactionin2) -- (nodein2) node[midway, above,font=\tiny] {$v_2$};
\draw[arrows = {-Stealth[length=8pt]}, dotted] (transactionink) -- (nodeink) node[midway, above,font=\tiny] {$v_k$};

\draw[arrows = {-Stealth[fill=none,length=8pt]}] (nodein1) -- (nodedelta) node[midway, above,font=\tiny] {$v_1$};
\draw[arrows = {-Stealth[fill=none,length=8pt]}] (nodein2) -- (nodedelta) node[midway, above,font=\tiny] {$v_2$};
\draw[arrows = {-Stealth[fill=none,length=8pt]}] (nodeink) -- (nodedelta) node[midway, above,font=\tiny] {$v_k$};

\draw[arrows = {-Stealth[fill=none,length=8pt]}, dotted] (nodeout1) -- (transactionout1) node[midway, above,font=\tiny] {$v_1^\prime$};
\draw[arrows = {-Stealth[fill=none,length=8pt]}, dotted] (nodeout2) -- (transactionout2) node[midway, above,font=\tiny] {$v_2^\prime$};
\draw[arrows = {-Stealth[fill=none,length=8pt]}, dotted] (nodeoutk) -- (transactionoutk) node[midway, above,font=\tiny] {$v_k^\prime$};

\draw[arrows = {-Stealth[length=8pt]}] (nodedelta) -- (nodeout1) node[midway, above,font=\tiny] {$v_1^\prime$};
\draw[arrows = {-Stealth[length=8pt]}] (nodedelta) -- (nodeout2) node[midway, above,font=\tiny] {$v_2^\prime$};
\draw[arrows = {-Stealth[length=8pt]}] (nodedelta) -- (nodeoutk) node[midway, above,font=\tiny] {$v_l^\prime$};

\end{tikzpicture}
}
\caption{Schematic of a transaction $\Delta$. Nodes with a single (resp. double) border symbolize TXOs (resp. transactions). TXOs consumed by $\Delta$ originate from prior transactions, while those created in $\Delta$ may serve as input TXOs in subsequent transactions.}
\label{fig:scheme_transaction}

\end{figure}

\paragraph{Transaction} A transaction $\Delta$ is defined by two subsets of TXOs: a set of input TXOs, $\Delta_\text{in}$ that will be fully consumed during the transaction, and a set of output TXOs, $\Delta_\text{out}$ that will be created during the transaction. A transaction can thus be seen as a transformation of TXOs that allows a change in the distribution of value among users represented by their scripts. A transaction output that has been consumed in a transaction does not exist anymore and thus cannot be double-spent in another transaction. A transaction output that has not been spent yet is called \textit{unspent transaction output} or UTXO. Let $v_\text{in}(\Delta) \triangleq \sum_{(\ket{p}, v) \in \Delta_\text{in}}v$ be the total value of the input TXOs and $v_\text{out}(\Delta) \triangleq \sum_{(\ket{p}, v) \in \Delta_\text{out}} v$ be the total value of the output TXOs.  For a transaction to be valid, it is imperative that $v_\text{in}(\Delta) - v_\text{out}(\Delta) \geq 0$ \footnote{This is not true for Coinbase transactions, but we will not consider them in this work.}.  We can see in Figure \ref{fig:scheme_transaction} a schematic representation of a transaction. The difference $v_\text{in}(\Delta) - v_\text{out}(\Delta) \geq 0$ is implicitly given to the miner as a form of remuneration for integrating the transaction into the chain.

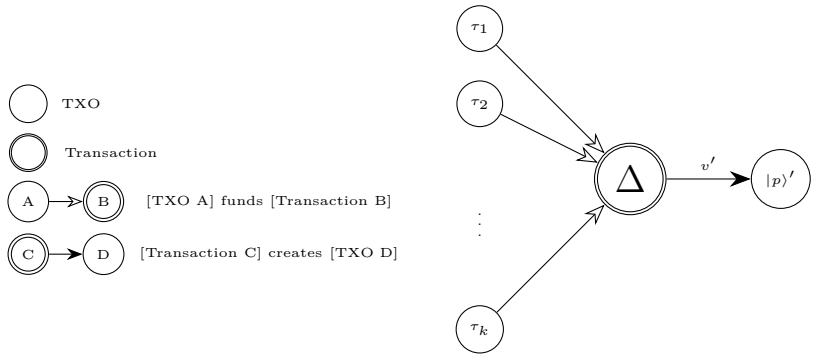
\begin{figure}[H]
\centering
\resizebox{0.9\textwidth}{!}{
\begin{tikzpicture}

\node[circle, draw, font=\tiny, minimum size=0.50cm] (A) at (-8, 1.) {};
\node[font=\tiny] (A) at (-7.3, 1.) {TXO};

\node[circle, draw, font=\tiny, minimum size=0.45cm] (A) at (-8, 0.35) {};
\node[circle, draw, font=\tiny, minimum size=0.50cm] (A) at (-8, 0.35) {};
\node[font=\tiny] (A) at (-6.9, 0.35) {Transaction};

\node[circle, draw, font=\tiny, minimum size=0.50cm] (TXOA) at (-8, -0.3) {A};
\node[circle, draw, font=\tiny, minimum size=0.50cm] (transactionB) at (-7, -0.3) {B};
\node[circle, draw, font=\tiny, minimum size=0.45cm] (A) at (-7, -0.3) {};
\draw[arrows = {-Stealth[length=6pt, fill=none]}] (TXOA) -- (transactionB);

\node[font=\tiny] (A) at (-4.8, -0.3) {[TXO A] funds [Transaction B]};

\node[circle, draw, font=\tiny, minimum size=0.50cm] (transationC) at (-8, -1.) {C};
\node[circle, draw, font=\tiny, minimum size=0.45cm] (A) at (-8, -1.) {};
\node[circle, draw, font=\tiny, minimum size=0.50cm] (TXOD) at (-7, -1.) {D};
\draw[arrows = {-Stealth[length=6pt]}] (transationC) -- (TXOD);

\node[font=\tiny] (A) at (-4.8, -1.) {[Transaction C] creates [TXO D]};

\node[circle, draw, font=\tiny] (nodein1) at (-2,2) {$\tau_1$};
\node[circle, draw, font=\tiny] (nodein2) at (-2,1) {$\tau_2$};
\node[circle, font=\tiny] (nodeinj) at (-2,-0.5) {$\vdots$};
\node[circle, draw, font=\tiny] (nodeink) at (-2,-2) {$\tau_k$};

\node[circle, draw, font=\Large, minimum size=0.95cm] (nodedelta) at (0,0) {};
\node[circle, draw, font=\Large, minimum size=0.8cm] (nodedelta1) at (0,0) {$\Delta$};

\node[circle, draw, font=\tiny] (nodeout1) at (2,0) {$\ket{p}^\prime$};


\draw[arrows = {-Stealth[fill=none,length=8pt]}] (nodein1) -- (nodedelta);
\draw[arrows = {-Stealth[fill=none,length=8pt]}] (nodein2) -- (nodedelta);
\draw[arrows = {-Stealth[fill=none,length=8pt]}] (nodeink) -- (nodedelta);

\draw[arrows = {-Stealth[length=8pt]}] (nodedelta) -- (nodeout1) node[midway, above,font=\tiny] {$v^\prime$};

\end{tikzpicture}
}
\caption{Schematic of a payment transaction $\Delta$.}
\label{fig:scheme_payment_transaction}

\end{figure}

\paragraph{Example of Payment Transaction} An illustrative example of a transaction is the transfer of payment from one user to another. Let us consider two users, $u$ and $u^\prime$, where $u$ intends to transfer a total value $v^\prime$ to $u^\prime$. Assuming that $u$ has control over a set of TXOs, denoted as $\tau_1, \tau_2, ..., \tau_k$ with a cumulative value exceeding $v$. The corresponding $\Delta$ takes the following form: $\Delta_\text{in} = \{ \tau_1, ..., \tau_k \}$ et $\Delta_\text{out} = \{ (\ket{p}^\prime, v^\prime) \}$, where $\ket{p}^\prime$ is a script belonging to $u^\prime$. This transaction is depicted in Figure \ref{fig:scheme_payment_transaction}.

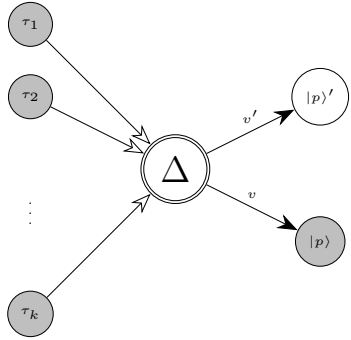
\begin{figure}[H]
\centering
\resizebox{0.4\textwidth}{!}{
\begin{tikzpicture}

\node[circle, draw, font=\tiny, fill=gray!50] (nodein1) at (-2,2) {$\tau_1$};
\node[circle, draw, font=\tiny, fill=gray!50] (nodein2) at (-2,1) {$\tau_2$};
\node[circle, font=\tiny] (nodeinj) at (-2,-0.5) {$\vdots$};
\node[circle, draw, font=\tiny, fill=gray!50] (nodeink) at (-2,-2) {$\tau_k$};

\node[circle, draw, font=\Large, minimum size=0.95cm] (nodedelta) at (0,0) {};
\node[circle, draw, font=\Large, minimum size=0.8cm] (nodedelta1) at (0,0) {$\Delta$};

\node[circle, draw, font=\tiny, fill=white!20] (nodeout1) at (2,1) {$\ket{p}^\prime$};
\node[circle, draw, font=\tiny, fill=gray!50] (nodeout2) at (2,-1) {$\ket{p}$};


\draw[arrows = {-Stealth[fill=none,length=8pt]}] (nodein1) -- (nodedelta);
\draw[arrows = {-Stealth[fill=none,length=8pt]}] (nodein2) -- (nodedelta);
\draw[arrows = {-Stealth[fill=none,length=8pt]}](nodeink) -- (nodedelta);

\draw[arrows = {-Stealth[length=8pt]}] (nodedelta) -- (nodeout1) node[midway, above,font=\tiny] {$v^\prime$};
\draw[arrows = {-Stealth[length=8pt]}] (nodedelta) -- (nodeout2) node[midway, above,font=\tiny] {$v$};

\end{tikzpicture}
}
\caption{Schematic of a payment transaction $\Delta$ with change. Gray (resp. white) TXOs belong to user $u$ (resp. $u^\prime$).}
\label{fig:scheme_payment_transaction_change}

\end{figure}

\paragraph{Change TXO} In the previous example, the portion of the input value that exceeds the payment value $v$ is given to the miner. This residual leftover can greatly exceed the standard miner reward, resulting in a loss for $u$. In order to receive back a part of the input value that is not used to fund the payment and to cover the usual miner fees, an additional output TXO is added to $\Delta_\text{out}$ to account for the change.  Therefore, we have $\Delta_\text{out} = \left\{ \left( \ket{p}^\prime, v^\prime \right), \left( \ket{p}, v \right) \right\}$ where $\ket{p}$ is an address of $u$. The miner will now receive $v_\text{in}(\Delta) - v - v^\prime \geq 0$. This transaction is depicted in Figure \ref{fig:scheme_payment_transaction_change}.

The change script is often referred to as the \textit{change script} or \textit{change address}. A script is said to be \textit{reused} if the script has been used to protect multiple TXOs. This is the case, for instance, if $u$ chooses to protect its change TXO with a script used to protect one of the input TXOs. However, to enhance privacy in Bitcoin, it is commonly recommended to use a new address for each change TXO to break the link between the change and the inputs.

\subsection{Script Clustering} 

Let $\mathcal{S}$ be a set of locking scripts. A \textit{clustering} of $\mathcal{S}$ is a partition of $\mathcal{S}$ consisting of non-empty and mutually exclusive subsets, such that their union equals $\mathcal{S}$. Each of these subsets is referred to as a \textit{cluster}. The \textit{number of clusters} is the number of subsets in the partition. Given a clustering $\mathcal{C}$ and a set of clusters $\mathcal{A} \subset \mathcal{C}$, we define the operation of \textit{merging} $\mathcal{A}$ as the process of replacing the current clustering $\mathcal{C}$ with a new clustering $\mathcal{C}^\prime$ such that $\mathcal{C}^\prime = (\mathcal{C} - \mathcal{A}) \cup \left\{ \bigcup_{c \in \mathcal{A}} c  \right\}$. This operation consolidates all clusters of $\mathcal{A}$ into a new, larger cluster. In the following, merging a subset of scripts $\mathcal{B}$ will be equivalent to merging all clusters in $\mathcal{C}$ that contain at least one script of $\mathcal{B}$, i.e. 
\begin{equation}
\{ c \in \mathcal{C} \ | \ c \cap \ \mathcal{B} \neq \emptyset \}
\end{equation}
   The \textit{atomic clustering} is the clustering consisting of all singletons, where each script forms its own cluster.

\section{Heuristics}

\label{sec:heuristics}

We reiterate that this paper aims to identify to identify scripts likely belonging to the same user or entity. Focusing solely on the entities rather than individual scripts will facilitate future analytical works. Uncovering entities equates to discerning a clustering among the scripts, where the resultant clusters will denote the final entities. To this end, we have to identify the sets of scripts to be merged. In this section, we will develop several heuristics that propose groups of scripts, which could reasonably belong to the same entity. These groups will thus be potential candidates for merging. These heuristics are based on information contained in the transaction's microstructure, particularly the management of TXOs and addresses. Subsequently, we will heavily employ two notations: $n_\text{in}(\Delta)$ and $n_\text{out}(\Delta)$, denoting the respective number of distinct input scripts and output scripts in $\Delta$.

\subsection{Common-Input-Ownership Heuristics}

A proposed transaction can be integrated into the chain if validated by the network. For a transaction $\Delta$ to be considered valid, each input TXO $(\ket{p}, v) \in \Delta_\text{in}$ is accompanied by an \textit{unlocking script} $\bra{q}$. This script, also written in Bitcoin script language, proves to the network that the owner of the mentioned TXO agrees to spend their TXO to fund $\Delta$. $\bra{q}$ usually contains one or more cryptographic signatures. These signatures encipher information using the private keys from which the identifiers in $\ket{p}$ are derived. These signatures are easily verifiable and tamper-proof. Finding a valid signature without knowledge of the private keys within a reasonable timeframe is highly unlikely. This way, the network can ensure that the owners of $(\ket{p}, v)$ indeed agree to spend their funds in $\Delta$. The responsibility of constructing a transaction, especially gathering all valid signatures, lies with the transaction initiator. If a single user owns all the input TXOs, the task is straightforward as he can calculate all the signatures using his private keys. Conversely, if the input funds belong to multiple owners, each owner will need to compute the signatures. This scenario requires collaboration and greater coordination efforts to construct the final transaction.  Heuristic \ref{heur:cio} often referred to as the \textit{common-input-ownership heuristic} states that it is more reasonable to assume that all scripts in $\Delta_\text{in}$ belong to the same user.

\begin{heur}[Common-Input-Ownership Heuristic] Let $\Delta$ be a  transaction, if: 
\label{heur:cio}
\begin{enumerate}
\item[a.] $n_\text{in}(\Delta) \geq 2$, 
\end{enumerate}
then the input scripts of $\Delta$ are merged.
\label{heur:common-input}
\end{heur}

However, it is not uncommon for users to coordinate in constructing a transaction, sometimes by engaging a third-party coordinator. This is precisely the case with CoinJoin transactions. These transactions pool funds from multiple users to mix them and challenge the heuristic \ref{heur:cio}. In a prior study \cite{schnoering2023heuristics}, the authors developed several heuristics to detect transactions that could reasonably be identified as CoinJoin. Utilizing these heuristics, we constructed a heuristic closely resembling heuristic \ref{heur:cio}, excluding transactions related to CoinJoin.

\begin{heur}[CoinJoin-resistant C-I-O Heuristic] Let $\Delta$ be a \newline transaction, if: 
\label{heur:ciocoinjoin}
\begin{enumerate}
\item[a.] $n_\text{in}(\Delta) \geq 2$, 
\item[b.] $\Delta$ is not a CoinJoin transaction (according to \cite{schnoering2023heuristics}),
\end{enumerate}
then the input scripts of $\Delta$ are merged.
\label{heur:common-input-coinjoin}
\end{heur}

\subsection{Change Address Heuristic}

\begin{figure}[H]
\centering
\resizebox{0.9\textwidth}{!}{
\begin{tikzpicture}


\node[circle, font=\tiny] (node0) at (-5,0) {};

\node[circle, draw, font=\tiny] (node1) at (-4,0) {$\ket{p}_1$};
\node[circle, draw, font=\tiny, minimum size=0.45cm] (transaction1) at (-3,0) {};
\node[circle, draw, font=\tiny, minimum size=0.4cm] (transaction11) at (-3,0) {};
\node[circle, font=\tiny] (node1bis) at (-2,1) {};

\node[circle, draw, font=\tiny] (node2) at (-2,0) {$\ket{p}_2$};
\node[circle, draw, font=\tiny, minimum size=0.45cm] (transaction2) at (-1,0) {};
\node[circle, draw, font=\tiny, minimum size=0.4cm] (transaction22) at (-1,0) {};
\node[circle, font=\tiny] (node2bis) at (0,1) {};

\node[circle, draw, font=\tiny] (node3) at (0,0) {$\ket{p}_3$};
\node[circle, draw, font=\tiny, minimum size=0.45cm] (transaction3) at (1,0) {};
\node[circle, draw, font=\tiny, minimum size=0.4cm] (transaction33) at (1,0) {};
\node[circle, font=\tiny] (node3bis) at (2,1) {};

\node[circle, draw, font=\tiny] (node4) at (2,0) {$\ket{p}_4$};
\node[circle, draw, font=\tiny, minimum size=0.45cm] (transaction4) at (3,0) {};
\node[circle, draw, font=\tiny, minimum size=0.4cm] (transaction44) at (3,0) {};
\node[circle, font=\tiny] (node4bis) at (4,1) {};

\node[circle, font=\tiny] (node5) at (4,0) {$\cdots$};

\draw[arrows = {-Stealth[length=4pt]}] (node0) -- (node1) node[midway, below,font=\tiny] {$v_0$};

\draw[arrows = {-Stealth[fill=none,length=4pt]}] (node1) -- (transaction1) {};
\draw[arrows = {-Stealth[length=4pt]}] (transaction1) -- (node2) node[midway, below,font=\tiny] {$v_1$};
\draw[arrows = {-Stealth[length=4pt]},  dotted] (transaction1) -- (node1bis) node[midway, above,font=\tiny] {$v_1^\prime$};

\draw[arrows = {-Stealth[fill=none,length=4pt]}] (node2) -- (transaction2) {};
\draw[arrows = {-Stealth[length=4pt]}] (transaction2) -- (node3) node[midway, below,font=\tiny] {$v_2$};
\draw[arrows = {-Stealth[length=4pt]},  dotted] (transaction2) -- (node2bis) node[midway, above,font=\tiny] {$v_2^\prime$};

\draw[arrows = {-Stealth[fill=none,length=4pt]}] (node3) -- (transaction3) {};
\draw[arrows = {-Stealth[length=4pt]}]  (transaction3)  -- (node4) node[midway, below,font=\tiny] {$v_3$};
\draw[arrows = {-Stealth[length=4pt]},  dotted] (transaction3)  -- (node3bis) node[midway, above,font=\tiny] {$v_3^\prime$};

\draw[arrows = {-Stealth[fill=none,length=4pt]}] (node4) -- (transaction4) {};
\draw[arrows = {-Stealth[length=4pt]}] (transaction4) -- (node5) node[midway, below,font=\tiny] {$v_4$};
\draw[arrows = {-Stealth[length=4pt]},  dotted]  (transaction4) -- (node4bis) node[midway, above,font=\tiny] {$v_4^\prime$};

\end{tikzpicture}
}
\caption{Schematic of a script chain. Full edges represent the transfer of change. Dotted edges represent payments.}

\label{fig:script_chain}

\end{figure}
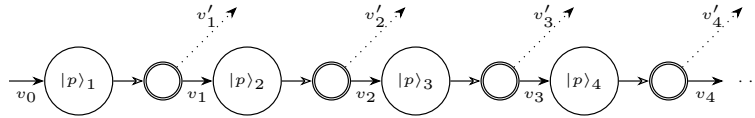

Fresh addresses / scripts are often created in order to protect the new (output) TXOs created by a transaction. Consider the simple payment transaction illustrated in figure \ref{fig:scheme_payment_transaction_change} as an example. Both users $u$ and $u^\prime$ can create a new address to receive the output TXOs. By generating a new script $\ket{p}^\prime$ to receive the payment of $u$, $u^\prime$ protects his privacy in two ways: hiding to $u$ his other addresses, and thus his other holdings, and preventing external users to track his funds. $u$ can also create a new address $\ket{p}$ to receive his change. By doing so, external observers will not be able to discern with certitude which address received the change and which address received the payment. If $u$ generates a new address whenever he makes a payment, his remaining funds are transferred to a new change address after each transaction. This creates a chain-like structure of scripts as depicted in figure \ref{fig:script_chain}. A \textit{script-chain} is defined as a sequence of scripts $\ket{p}_1 \rightarrow \ket{p}_2 \rightarrow ... \rightarrow \ket{p}_k$ satifying:
\begin{itemize}
\item all scripts belong to the same user $u$, 
\item each script is used only once,
\item $\ket{p}_{i+1}$ protects the change TXO of a transaction funded by $\ket{p}_i $ 
\end{itemize}
Only the most recent script (the leaf) in the chain holds the funds belonging to $u$.

If $u$ uses a HD wallet, he is very likely to use fresh change addresses after each transaction. If the recipient of $u$'s payment uses a fresh payment address then it is not possible to distinguish the change address among the recipient scripts. However, if the recipient reuses an address, then it is possible to infer that the other output address is the change address. Heuristic \ref{heur:change-address} analyses single-input payment transaction (conditions a. and b.) to detect if - the input script and one output script are likely to be derived from a HD wallet (conditions c. and d.), and - the other script has been reused (condition e.). 

\begin{heur}[Detection of Change Addresses] Let $\Delta$ be a transaction, if: 
\begin{enumerate}
\item[a.] $\left| \Delta_\text{in} \right| = 1$, the input script will be denoted $\ket{p}_\text{in}$,
\item[b.] $\left| \Delta_\text{out} \right| =   n_\text{out}(\Delta) = 2$,
\item[c.] $\ket{p}_\text{in}$ has not been reused (according to available information),
\item[d.] there exists exactly one output script $\ket{p}_\text{change}$ that has not been reused (according to available information),
\item[d.] the other output script $\ket{p}_\text{pay}$ has been reused (according to available information),
\end{enumerate}
then $\ket{p}_\text{in}$ and $\ket{p}_\text{change}$ are merged.
\label{heur:change-address}
\end{heur}

It is specified 'according to available information' in certain conditions, as determining whether a script has been reused or not depends on the other transactions known to us. For example, if an address $\ket{p}$ is reused for the first time at date $t^\prime$ but we are only studying transactions before date $t < t^\prime$, then this knowledge is unavailable to us. This heuristic can be compared to other widely used 'change heuristics'  \cite{androulaki2013evaluating,meiklejohn2013fistful,zhang2020heuristic} (see appendix \ref{app:heuristics}).

\subsection{Round Output Value Heuristic}

It is a well-known fact that humans exhibit a preference for round numbers. This phenomenon has been studied through the analysis of transaction amounts within a large database of mobile transactions \cite{wang2023last}. If we assume that the same psychological bias influences Bitcoin transactions, payments transactions designed by humans are likely to exhibit rounded amounts. Consequently, the payment output  is expected to have a rounded value (in satoshis), specifically a multiple of $10^i$. Conversely, the mining fee ($v_\text{in}(\Delta) - v_\text{out}(\Delta)$) is dependent on the transaction size and network congestion, and is therefore calculated algorithmically. As the mining fees are deducted from the change output, the value of the change output is likewise not subject to the rounding bias. Since users frequently think in terms of dollars, the rounding exponent $i$ is likely to fluctuate with the price of a satoshi in dollars. Let $p$ be the price of a satoshi in dollars, and let $x$ be a 'small' value in dollars in comparison to typical payment value. We choose $i$ as the largest integer satisfying $10^i \times p \leq x$, i.e. the integer part of $\log_{10} \left( \frac{x}{p} \right)$. Let $y \gg x$ represent the amount sent, $y$ is likely to be a multiple of $10^i$ satoshis due to the rounding bias. Heuristic \ref{heur:round-output} analyzes payment transaction (conditions a. and b.) to detect if - an output has a round amount and is not reused (condition d.), and - the other output amount is rounded to a lesser precision (condition e).

\begin{heur}[Detection of Round Output Value] Let $\Delta$ \newline be a transaction, $p$ be the price of a satoshi at the time of the transaction, and $x$ be a 'small' amount in dollars. If:
\begin{enumerate}
\item[a.] $\left| \Delta_\text{in} \right| = 1$, the input script will be denoted $\ket{p}_\text{in}$,
\item[b.] $\left| \Delta_\text{out} \right| =   n_\text{out}(\Delta) = 2$,
\item[c.] $\ket{p}_\text{in}$ has not been reused (according to available information),
\item[d.] there exists one output script $\ket{p}_\text{pay}$ that has not been reused (according to available information), and whose value is a multiple of $10^i$, with $i$ the integer part of $\log_{10} \left( \frac{x}{p} \right)$,
\item[e.] the value of the other output script is not a multiple of $10^{i - j}$, with $0 \leq j < i$,
\end{enumerate}
then $\ket{p}_\text{in}$ and $\ket{p}_\text{change}$ are merged.
\label{heur:round-output}
\end{heur}

\begin{figure}[H]
  \centering
  \resizebox{0.6\textwidth}{!}{
    \begin{tikzpicture}
    \footnotesize
      \begin{axis}[
        xlabel={Block Index ($k$)},
        ylabel={Rounding exponent $i$},
        grid=major,
        legend columns = 2,
        legend style={at={(0.5, 1.2)},anchor=north,nodes={scale=0.65, transform shape}}
      ]
      
      \addplot[] coordinates {
        (160000, 7)
        (192000, 7)
        (192000, 6)
        (229000, 6)
        (229000, 5)
        (231000, 5)
        (231000, 6)
        (232000, 6)
        (232000, 5)
        (244000, 5)
        (244000, 6)
        (249000, 6)
        (249000, 5)
        (446000, 5)
        (446000, 4)
        (447000, 4)
        (447000, 5)
        (453000, 5)
        (453000, 4)
        (497000, 4)
        (497000, 3)
        (507000, 3)
        (507000, 4)
        (510000, 4)
        (510000, 3)
        (513000, 3)
        (513000, 4)
        (582000,4)
        (582000,3)
        (587000, 3)
        (587000, 4)
        (588000, 4)
        (588000, 3)
        (592000, 3)
        (592000, 4)
        (593000, 4)
        (593000, 3)
        (596000, 3)
        (596000, 4)
        (617000, 4)
        (617000, 3)
        (619000, 3)
        (619000, 4)
        (641000, 4)
        (641000, 3)
        (700000, 3)
      };

      \end{axis}
     
    \end{tikzpicture}
  }
  \caption{Evolution of the rounding exponent $i$ for $x=1$ dollar w.r.t. the block index. }
  \label{fig:rounding_exp}
\end{figure}
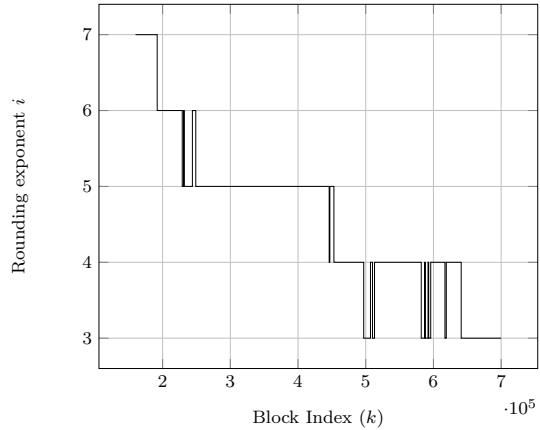

We plotted on Figure \ref{fig:rounding_exp} the evolution of the rounding exponent as a function of the price of a satoshi and a size of $x = 1$ dollar. To calculate the price of a satoshi, we retrieved the Bitcoin price in dollars from \url{blockchain.com}; our historical data goes back to January 1, 2012.

\subsection{Force Merge of Inputs Heuristic}

If the user $u$ uses a HD wallet, he is very likely to generate a new script whenever he receives a payment. Consequently, each new reception address is possibly the root of a new script-chain. Thus, $u$ may have several script-chains under his control. If $u$ intends to transfer funds, denoted as $v$, to another user $u^\prime$, he utilizes the unspent transaction outputs (UTXOs) protected by the leaves of his script-chains. To prioritize the preservation of his privacy, he attempts to spend funds from a single leaf /script, thereby avoiding having his scripts appear together in a same transaction. However, if no individual chain possesses enough funds to cover the transaction cost, $u$ is compelled to consolidate UTXOs from multiple chains to fund the transaction. Consequently, this consolidation of transaction outputs establishes a  link between several of scripts belonging to $u$. We leverage this observation to develop heuristic \ref{heur:force_merge} for clustering scripts likely to be owned by the same user. Heuristic \ref{heur:force_merge} analyzes multi-input payment transactions (conditions a. and b.) to detect if - the input scripts and the supposed change script are likely to be derived from a HD wallet (conditions c. and d.), and - the privacy leakage is minimized, i.e. the set of input TXOs merged to reach $v$ is minimal (condition e.).

\begin{heur}[Detection of Forced Merge of Inputs] Let $\Delta$ be a transaction, if:
\begin{enumerate}
\item[a.] $\left| \Delta_\text{in} \right| = n_\text{in}(\Delta) \geq 2$, 
\item[b.] $\left| \Delta_\text{out} \right| = n_\text{out}(\Delta) =  2$, we suppose that the output script associated with the higher value $v_\text{max}$, denoted $\ket{p}_\text{max}$  is the payment output while the other script $\ket{p}_\text{min}$ is the change script.  
\item[c.] No input script has been reused (according to available information)
\item[d.] The change script has not been reused (according to available information)
\item[e.] $v_\text{in}(\Delta) - \min_{(v, \ket{p}) \in \Delta_\text{in}}v \ < \ v_\text{max}$
\end{enumerate}
then the input scripts and $\ket{p}_\text{min}$ are merged. 
\label{heur:force_merge}
\end{heur}

\subsection{Service Deposit Address Heuristic}

Since the emergence of Bitcoin, a wide range of services has been developed, encompassing exchanges, gambling platforms, marketplaces, and more. To utilize these services, users are required to deposit their funds into addresses associated with the specific service. Due to the ease of creating new addresses, these services often have the capacity to generate dedicated deposit addresses for individual customers. Consequently, tracking the deposited funds and determining which funds are available for each customer becomes relatively straightforward. A typical service may exercise control over thousands of addresses, with the largest ones even managing millions. Subsequently, the funds deposited into these addresses are periodically transferred to specialized wallets known as hot or cold wallets, where they are more securely stored. Hot wallets refer to  wallets that are connected to the network, enabling easy access and quick transactions, these addresses are in general used to honor customer withdrawals. Cold wallets are offline storage devices, designed to enhance security by keeping funds of the service disconnected from online threats. To save transaction fees, the funds from a  large set of  deposit addresses are moved together in the same transaction, we use the heuristic \ref{heuristic:service_deposit} to detect such consolidation transactions (many inputs to one output). The heuristic is a special case of the heuristic \ref{heur:cio}.

\begin{heur}[Detection of Service Deposit Addresses] Let $\Delta$ be a transaction, if: 
\begin{enumerate}
\item[a.] $n_\text{in}(\Delta) \geq a$, 
\item[b.] $n_\text{out}(\Delta) = 1$, 
\end{enumerate}
\label{heuristic:service_deposit}
then the input scripts of $\Delta$ belong to the same user. $a$ is an input parameter. 
\end{heur}

\section{Results}
\label{sec:results}

This section aims to assess the effectiveness of the clustering heuristics developed previously. For each heuristic $h$, we iteratively construct clusters by applying $h$ to each transaction and merging scripts associated with the same entity as determined by $h$.  On the blockchain, transactions are grouped by blocks, and these blocks form a chain to which a new block of transactions is added approximately every 10 minutes. The term 'block index' refers to a block's position within the chain. Let $k$ be a block index, we define as $\mathcal{S}_k$ the set of scripts present in transactions up to the block indexed by $k$. Using a clustering heuristic $h$, an iterative clustering construction can be performed as described below:

\begin{itemize}
\item Start from the trivial clustering.  
\item For every block $k$, analyze every transaction and consolidate scripts identified as belonging to the same entity as per heuristic $h$. Continue processing subsequent blocks with the updated clustering.
\end{itemize}

We denote by $\mathcal{C}_k^h$ the clustering obtained with heuristic $h$ by processing every block up to index $k$. 

\paragraph{Clustering ratio} In order to evaluate the clustering effectiveness of $h$, we compare the final number of entities to the initial number of entities. In our case,  the initial and final number of entities are respectively the number of scripts $|\mathcal{S}_k|$ and the number of clusters $| \mathcal{C}_k^h|$. We define the \textit{clustering ratio} as follows: 

\begin{equation}
r^h_k = \frac{|\mathcal{C}_k^h|}{|\mathcal{S}_k|} \in (0, 1]
\end{equation}

The closer this number is to 0, the higher the clustering effectiveness of $h$.

\begin{figure}[H]
  \centering
  \resizebox{0.6\textwidth}{!}{
    \begin{tikzpicture}
    \footnotesize
      \begin{axis}[
        xlabel={Block Index ($k$)},
        ylabel={Number of Scripts Observed ($\mathcal{S}_k$)},
        ymode=log
      ]
      
      \addplot[mark=none] coordinates {
        (100000, 174000)
        (200000, 660000)
        (300000, 35600000)
        (400000, 129300000)
        (500000, 346800000)
        (600000, 569100000)
        (700000, 874600000)
      };
      
      \end{axis}
    \end{tikzpicture}
  }
  \caption{Evolution of the number of scripts w.r.t. the block index. }
  \label{fig:num_scripts}
\end{figure}
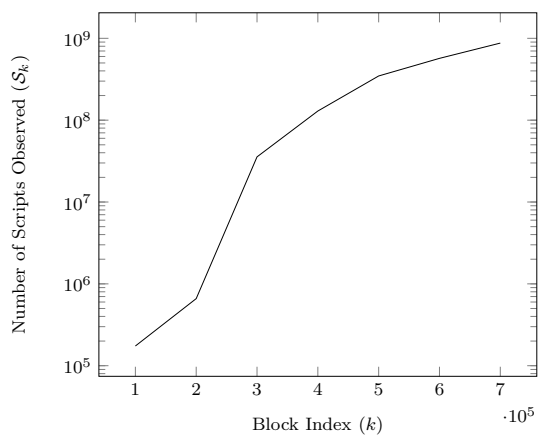

\paragraph{Data} To extract data from the Bitcoin blockchain, we have set up a Bitcoin Core full node. This required downloading and syncing the complete transaction ledger from a network of peers. After installing the latest Bitcoin Core software and configuring the node, the entire transaction history was saved in the local blockchain data directory, specifically in the 'blkXXXXX.dat' files in the '/.blocks' folder. We then delved into the Bitcoin protocol and file structure, using parsing techniques to extract transaction details. This process ensured accurate data for our analysis. We have reported in table \ref{tab:num_scripts} the total number of scripts in $\mathcal{S}_k$ for different values of $k$. We have also plotted in figure \ref{fig:num_scripts} the evolution of the total number of scripts w.r.t the block index. It can be observed that the number of scripts has exponentially increased over the years.

\begin{table}[H]
\footnotesize
    \centering
    \begin{tabular}{|c|c|}
        \hline
        \textbf{Block Index ($k$)} & \textbf{Number of Scripts Observed ($\mathcal{S}_k$)}  \\
        \hline
        100000 & 174K  \\
        \hline
        200000 & 6.6M  \\
        \hline
        300000 & 35.6M  \\
        \hline
        400000 & 129.3M  \\
        \hline
        500000 & 346.8M  \\
        \hline
        600000 & 569.1M  \\
        \hline
        700000 & 874.6M  \\
        \hline
    \end{tabular}
    \caption{Total number of scripts.}
    \label{tab:num_scripts}
\end{table}

\paragraph{Results} 

\begin{figure}
  \centering
  \resizebox{0.99\textwidth}{!}{
    \begin{tikzpicture}
    \footnotesize
      \begin{axis}[
      ybar,
        xlabel={Block Index ($k$)},
        ylabel={Clustering ratio ( $r^h_k$ in \%)},
        symbolic x coords = {100000, 200000, 300000, 400000, 500000, 600000, 700000},
        legend columns = 2,
        ymax=105,
        label style={font=\tiny},
        tick label style={font=\tiny},
        nodes near coords,
        every node near coord/.append style={font=\tiny, rotate=90, anchor=west},
        bar width=5,
        legend style={at={(0.5, 1.2)},anchor=north,nodes={scale=0.65, transform shape}}
      ]

       \addplot[fill=black!60!white] coordinates {
        (200000, 98)
        (300000, 96.4)
        (400000, 97.1)
        (500000, 94)
        (600000, 93.9)
        (700000, 94.9)
      };
      \addlegendentry{Round Output Value ($x=1$, $j=1$)}
      
      \addplot[pattern=north east lines] coordinates {
        (100000, 96.9)
        (200000, 94.6)
        (300000, 96.1)
        (400000, 96.3)
        (500000, 94.6)
        (600000, 94.4)
        (700000, 94.5)
      };
      \addlegendentry{Force Merge of Inputs}
      
          \addplot[] coordinates {
   	(100000, 83.6)
        (200000, 70.9)
        (300000, 82.8)
        (400000, 87.5)
        (500000, 86)
        (600000, 86)
        (700000, 86.4)
      };
      \addlegendentry{Change Address}

      \end{axis}
     
    \end{tikzpicture}
  }
  \caption{Evolution of the clustering ratio for the change heuristic and its derivatives. }
  \label{fig:clustering_ratio_diff_heuristics2}
\end{figure}
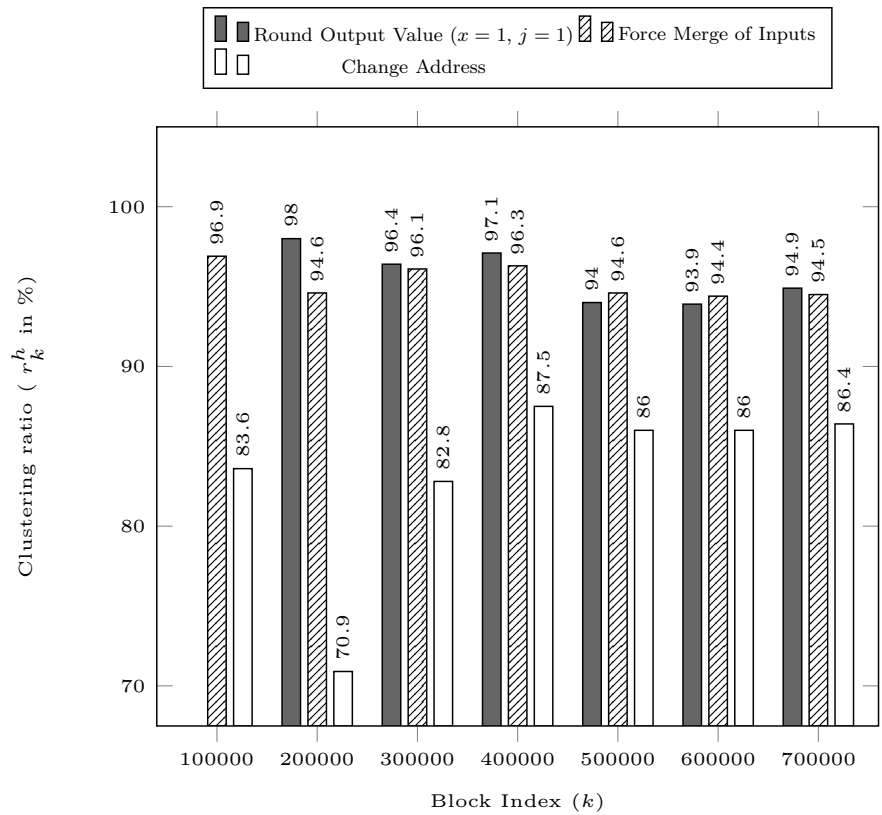

The clustering ratios for the different heuristics developed in the preceding section were computed across various block indices.  In implementing the Change Address and Force Merge of Inputs heuristics, we considered transaction information up to block index 700,000. When applying the Round Output Value heuristic, we fixed $x$ at $1$ dollar and $j$ at $1$. For the Service Deposit Address heuristic, we set $a$ to $25$. The evolution of clustering ratios relative to block index for various heuristics is illustrated in Figures \ref{fig:clustering_ratio_diff_heuristics} and \ref{fig:clustering_ratio_diff_heuristics2}. Excluding Bitcoin's first years of existence, we observe relatively stable clustering ratios from block 400,000 onwards. The C-I-O-type heuristics demonstrate the highest efficiency, effectively halving the number of entities that need to be studied. At block index 700,000, the discrepancy in clustering ratios between the two C-I-O heuristics is approximately 0.7\%, resulting in a difference of 6 million clusters between their respective final clusterings. The Change Address heuristic is the second most efficient, diminishing the number of scripts by approximately 15\%. Finally, the three remaining heuristics (Round Output Value, Force Merge of Inputs, and Deposit Addresses) each decrease the number of scripts by 5\% to 10\%.

\begin{figure}
  \centering
  \resizebox{0.99\textwidth}{!}{
    \begin{tikzpicture}
    \footnotesize
      \begin{axis}[
      ybar,
        xlabel={Block Index ($k$)},
        ylabel={Clustering ratio ( $r^h_k$ in \%)},
        symbolic x coords = {100000, 200000, 300000, 400000, 500000, 600000, 700000},
        legend columns = 2,
        ymax=105,
        label style={font=\tiny},
        tick label style={font=\tiny},
        nodes near coords,
        every node near coord/.append style={font=\tiny, rotate=90, anchor=west},
        bar width=5,
        legend style={at={(0.5, 1.2)},anchor=north,nodes={scale=0.65, transform shape}}
      ]

      \addplot[] coordinates {
        (100000, 71.5)
        (200000, 57.9)
        (300000, 46.4)
        (400000, 41.7)
        (500000, 49)
        (600000, 48.4)
        (700000, 46.8)
      };
      \addlegendentry{C-I-O}
      
       \addplot[fill=black!60!white] coordinates {
        (100000, 71.5)
        (200000, 58)
        (300000, 46.8)
        (400000, 42.4)
        (500000, 49.5)
        (600000, 48.8)
        (700000, 47.5)
      };
      \addlegendentry{CoinJoin-resistant C-I-O}

          \addplot[pattern=north east lines] coordinates {
        (100000, 90.4)
        (200000, 98)
        (300000, 96.9)
        (400000, 96.6)
        (500000, 93)
        (600000, 90.5)
        (700000, 88.5)
      };
      \addlegendentry{Deposit Address ($a=25$)}

      \end{axis}
     
    \end{tikzpicture}
  }
  \caption{Evolution of the clustering ratio for the C-I-O heuristic and its derivatives. }
  \label{fig:clustering_ratio_diff_heuristics}
\end{figure}
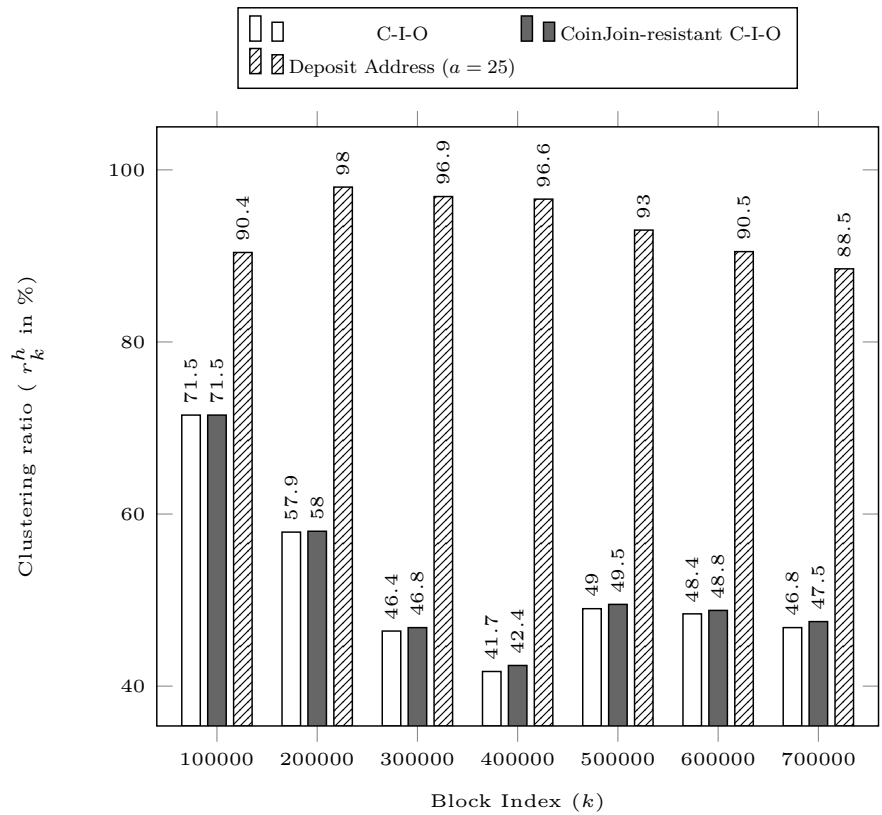

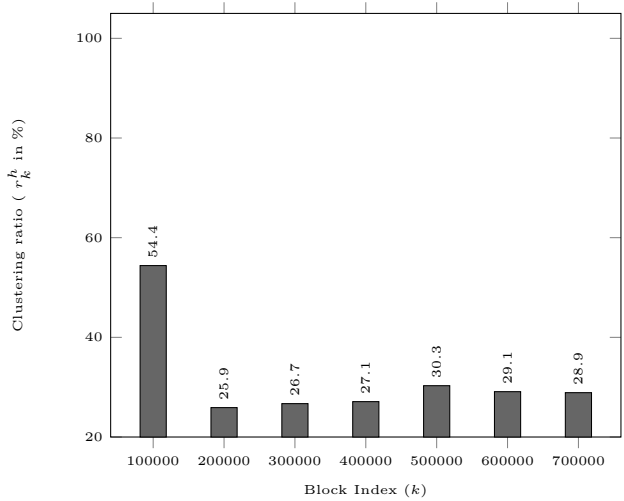
\begin{figure}[H]
  \centering
  \resizebox{0.7\textwidth}{!}{
    \begin{tikzpicture}
    \footnotesize
      \begin{axis}[
      ybar,
      ymin=20,
      ymax=100,
        symbolic x coords = {100000, 200000, 300000, 400000, 500000, 600000, 700000},
        legend columns = 2,
        ymax=105,
        label style={font=\tiny},
        tick label style={font=\tiny},
        nodes near coords,
        every node near coord/.append style={font=\tiny, rotate=90, anchor=west},
        xlabel={Block Index ($k$)},
        ylabel={Clustering ratio ( $r^h_k$ in \%)},
      ]
      
      \addplot[fill=black!60!white] coordinates {
        (100000, 54.4)
        (200000, 25.9)
        (300000, 26.7)
        (400000, 27.1)
        (500000, 30.3)
        (600000, 29.1)
        (700000, 28.9)
      };

      \end{axis}
     
    \end{tikzpicture}
  }
  \caption{Evolution of the clustering ratio for the Combined Heuristic w.r.t. the block index. }
  \label{fig:clustering_combined}
\end{figure}

\paragraph{Discussion} The impressive performance of the C-I-O heuristic largely explains why most studies utilize and rely on it to reduce the number of entities under scrutiny. For studies that do not want to merge entities involved in CoinJoin transactions, the CoinJoin-resistant C-I-O achieves nearly comparable performance. The Service Deposit Address heuristic is a specific case of the C-I-O heuristic, explaining its lower performance. However, the increasing clustering power of the Service Deposit Address may indicate widespread adoption of deposit addresses by services. The Change Address heuristic and its derivatives exhibit lesser reduction performance. The observed clustering stability with the Change Address and Force Merge of Inputs heuristics may suggest widespread adoption of fresh change addresses, driven by the increasing use of HD wallets since February 2012 (around block index 170,000). In addition to lower performance, these heuristics rely on stronger assumptions and require more resources, especially to determine address reuse. It is not surprising, therefore, that these heuristics are generally less utilized. However, they do identify addresses belonging to the same entity under conditions other than the C-I-O heuristic. If the reduction performance of the C-I-O heuristic is insufficient, combining these heuristics with a C-I-O-type heuristic should further decrease the number of entities and facilitate analysis. We thus combined four heuristics: CoinJoin-resistant C-I-O, Change Address, Round Output Value, and Force Merge of Inputs, aiming to capitalize on the individual clustering power of each heuristic. The evolution of the clustering ratio achieved through this Combined Heuristic is illustrated in Figure \ref{fig:clustering_combined}. This ratio remains relatively stable over blocks, reaching a 70\% reduction. This implies that the initial study of 874 million entities at block 700,000 is reduced to approximately 250 million clusters, representing a reduction of 624 million entities.

\newpage
\section*{Conclusion}

In this study, we explored various heuristics aimed at inferring groups of addresses likely belonging to the same entity. The application of one or more heuristics allows for clustering addresses on the Bitcoin blockchain into groups belonging to the same user. Consequently, an analytical study can reasonably choose to investigate these groups of entities rather than the initial set of addresses. Moreover, this clustering step reduces the number of entities to study, thus saving resources and facilitating the use of algorithms with high computational complexity. We introduced six heuristics, providing motivation by explaining their underlying assumptions and the behaviors they aim to detect. We then measured their reduction power by comparing their initial number of entities to the final number of clusters through the clustering ratio. For each heuristic, we also examined the evolution of this ratio over time to assess the efficiency of these heuristics. The common-input heuristic, almost universally employed in analytical work, halves the number of addresses. Other heuristics exhibit lower reduction powers, ranging from 5\% to 15\%. However, combinations of these heuristics with the common-input-ownership heuristic further enhance their effectiveness. This is explained by the fact that these heuristics were designed to detect different patterns of behavior or transactions. A strategic combination of a common-input heuristic with a change address heuristic, along with two other heuristics, achieves a 70\% reduction in the number of addresses. This translates to the study of 250 million clusters instead of 874 million addresses at block 700,000.

\newpage

\bibliographystyle{unsrt}
\bibliography{biblio}

\newpage

\begin{appendices}

\section{Heuristics from Related Works}
\label{app:heuristics}

\begin{heur}[Shadow Address \cite{androulaki2013evaluating}] Let $\Delta$ be transaction, if: 
\begin{enumerate}
\item[a.] $\left| \Delta_\text{out} \right| =   n_\text{out}(\Delta) = 2$ (two output scripts / addresses),
\item[b.] there exists one output script $\ket{p}_\text{change}$ that has not been used before,
\item[c.] the other output script $\ket{p}_\text{pay}$ has already been used before
\end{enumerate}
then $\ket{p}_\text{change}$ is the shadow (change) address. Thus, the scripts of $\Delta_\text{in}$ and $\ket{p}_\text{change}$ are merged.
\label{heur:change-address-androulaki}
\end{heur}

\begin{heur}[One-time Change Addresse \cite{meiklejohn2013fistful}] Let $\Delta$ be transaction, if: 
\begin{enumerate}
\item[a.] the set of input scripts (i.e. present in $\Delta_\text{in}$) does not intersect the set of output scrips (i.e. present in $\Delta_\text{out}$) (no self-change), 
\item[b.] there exists exactly one output script $\ket{p}_\text{change}$ that has not been used before,
\end{enumerate}
then $\ket{p}_\text{change}$ is the change address. Thus, the scripts of $\Delta_\text{in}$ and $\ket{p}_\text{change}$ are merged.
\label{heur:change-address-meiklejohn}
\end{heur}

\begin{heur}[Address reused-based Change Address \cite{zhang2020heuristic}] Let $\Delta$ be transaction, if: 
\begin{enumerate}
\item[a.] the set of input scripts (i.e. present in $\Delta_\text{in}$) does not intersect the set of output scrips (i.e. present in $\Delta_\text{out}$) (no self-change), 
\item[b.] there exists exactly one output script $\ket{p}_\text{change}$ that has not been used before and that is not reused after,
\end{enumerate}
then $\ket{p}_\text{change}$ is the change address. Thus, the scripts of $\Delta_\text{in}$ and $\ket{p}_\text{change}$ are merged.
\label{heur:change-address-zhang}
\end{heur}

\end{appendices}

\end{document}